\documentclass[aps,twoside,twocolumn]{revtex4}
\usepackage{epsfig}
\usepackage{amsmath,amssymb,color}
\usepackage[english]{babel}

\parskip=\medskipamount

\newcommand{\eq}[1]{(\ref{#1})}
\newcommand{\fig}[1]{Fig.\ref{#1}}
\newcommand{\be}{\begin{equation}}
\newcommand{\ee}{\end{equation}}
\newcommand\disp{\displaystyle}

\begin{document}

\title{Chaotic Hamiltonian systems revisited: Survival probability}

\author{V.A. Avetisov$^{1}$, S.K. Nechaev$^{2,3,4}$}

\affiliation{$^1$ N.N. Semenov Institute of Chemical Physics of the Russian Academy of
Sciences, 1199911, Moscow, Russia \\ $^2$LPTMS, Universit\'e Paris Sud, 91405 Orsay Cedex, France \\
$^3$ P.N. Lebedev Physical Institute of the Russian Academy of Sciences, 119991, Moscow, Russia \\
$^4$ J.-V. Poncelet Labotatory, Independent University, 119002, Moscow, Russia}

\date{\today}

\begin{abstract}

We consider the dynamical system described by the area--preserving standard mapping. It is known
for this system that $P(t)$, the normalized number of recurrences staying in some given domain of
the phase space at time $t$ (so-clled "survival probability") has the power--law asymptotics,
$P(t)\sim t^{-\nu}$. We present new semi--phenomenological arguments which enable us to map the
dynamical system near the chaos border onto the effective "ultrametric diffusion" on the boundary of
a tree--like space with hierarchically organized transition rates. In the frameworks of our
approach we have estimated the exponent $\nu$ as $\nu=\ln 2/\ln (1+r_g)\approx 1.44$, where
$r_g=(\sqrt{5}-1)/2$ is the critical rotation number.

\end{abstract}

\maketitle

\section{Introduction}

In this paper we propose a new estimate for the "Poincar\'e recurrences" (or "survival
probability"), $P(t)$, for the standard mapping in the vicinity of the chaos border
\cite{karney1,chir1}. Our consideration explicitly exploits the fact that the phase space of the
standard mapping near the border separating chaotic and integrable behaviors has self--similar
scale--invariant structure consisting of hierarchical set of metastable islands of integrability
("cantori") embedded in the "chaotic sea" \cite{mackay}. To be specific, we consider the
area--preserving standard mapping
\be
\begin{array}{l}
y_{t+1} = y_t - K/(2\pi) \sin 2\pi x_t \medskip \\
x_{t+1} = x_t + y_{t+1} \quad \mod 1
\label{eq:rec}
\end{array}
\ee
For $K\in [0, K_g[$ the phase space of the system has disjoined islands of integrability which is
destroyed as $K\to K_g$ from below, where $K_g\approx 0.97163540631$ \cite{mackay}. Above the
critical value $K_g$ the behavior of the system is less universal: many invariant
Kolmogorov--Arnold--Moser (KAM) tori disappear, but still some islands of metastability survive
around the biggest resonances. However, we should emphasize that the reorganization of the phase
space above the value $K_g$ has no consequence for our consideration since we do not touch the
region $K>K_g$ and are interested in the survival probability only when $K$ approaches $K_g$ from
below.

Our consideration of the survival probability is semi--phenomenological, that is we relay only on
the measurable "macroscopic" characteristics acquired in course of the iteration of the map
\eq{eq:rec}. To be precise, we rely on the following well--established and confirmed facts: i) The
number of principal resonances follows the Fibonacci sequence when $K\nearrow K_g$
\cite{hanson,chir2}; ii) The generic behavior of phase trajectories is as follows: the phase
trajectory stays in the vicinity of some resonance (low--flux cantori) and then rapidly crosses the
chaotic sea until another metastable low--flux cantori is reached \cite{hanson,mackay2}; iii) The
phase space of the standard mapping is self--similar being usually represented by a binary (i.e.
3--branching) Cayley tree \cite{hanson,meiss1,meiss2}; iv) The survival probability has power--law
asymptotic behavior, $P(t)\sim t^{-\nu}$ (for the first time this has been shown in
\cite{karney1,chir1}).

Remind that survival probability is the normalized number of recurrences \eq{eq:rec} which stay in
some given domain of the phase space at time $t$. Different research groups present different
arguments for estimates of $\nu$, typically $1<\nu \lesssim 3$. The most intriguing contradiction
concerns the discrepancy in reported values of $\nu$. The numerical simulations \cite{chir1,chir3}
demonstrate $\nu\approx 1.4 \div 1.5$, while almost all known analytic constructions give
essentially larger exponents: $\nu\approx 1.96$ in \cite{meiss1} and $\nu\approx 3.05$ in
\cite{meiss2}. The scaling analysis \cite{chir2} valid just near the chaos boundary gives $\nu=3$.
The special attention should be paid to the recent works \cite{ketzm,ven1,ven2}. In \cite{ketzm}
the authors demonstrate that by an appropriate randomization of the "Markov tree model" proposed in
\cite{meiss1,meiss2} one can arrive at the value $\nu\approx 1.57$. The works \cite{ven1,ven2}
claim $\nu=3$ for sticking of trajectories near $K_g$ in agreement with \cite{chir2} and $\nu=3/2$
for trapping of chaotic trajectories in the vicinity of cantori for $K\approx 2\ell/\pi$ (where
$\ell$ is a nonzero integer). The similar exponent, $\nu=3/2$, was also obtained for a standard map
in the work \cite{verg}. There is a point of view that the value $\nu\approx 1.4 \div 1.5$
corresponds to an intermediate behavior of the system which is not yet reached the stationary
regime -- see the corresponding discussion in \cite{chir2,weiss,chir4,art}.

The main aim of the present letter is to present some simple arguments in favor of the statement
that the value $\nu\approx 1.4 \div 1.5$ could be the actual decay exponent of the survival
probability $P(t)$ for $t\to\infty$ without additional randomization of local transitional
probabilities. Let us emphasize once more that in our approach we get rid of microscopic
consideration of the detailed structure of quasiperiodic orbits and corresponding fluxes, but
characterize the phase space of our system for a given coupling constant, $K$ in the vicinity of
the critical value $K_g$, just by the hierarchy of resonances.

According to \cite{greene,mackay,chir5}, the structure of the critical KAM curve is determined by
arithmetic properties of the rotation number, $r$, in the continued fraction representation:
\be
r=\frac{1}{\disp m_1+\frac{1}{\disp m_2+\frac{1}{\disp m_3+...}}}=[m_1\,m_2\,m_3\,...]
\label{eq:0}
\ee
The $n$s best convergent to the rotation number $r$ is $r_n=[m_1\,m_2\,...\,m_n]=p_n/q_n$ with
$m_1=m_2=...=m_n=1$, where $q_n$ is the Fibonacci number. Recall that the Fibonacci numbers
satisfy the recursion relation $q_{n+1}=q_n+q_{n-1}$; $q_0=0, q_1=1$.

For $K \nearrow K_g$ the periodic trajectories with rotation numbers $r_n$ determine the structure
of the phase space, converging to the critical boundary curve with $r_g$ (see \cite{mackay}). In
the limit $n\to\infty$ (i.e. for $q_n\to\infty$) the phase space becomes self--similar with the
scaling factor $s_n=q_n/q_{n-1}\to 1+r_g\approx 1.618$, where $r_g=[111...]=(\sqrt{5}-1)/2$ is the
"golden mean". The convergents $r_n$ ($n$ is fixed) characterize the positions of unstable fixed
points of resonances for a given value of a coupling constant, $K$.

According to \cite{chir2}, the average local exit time, $\tau_n$, from a given scale $n$ can be
estimated as $\tau_n\sim |r_g-r'_n|^2/D_n$, where $r'_n$ is the $n$s convergent of the critical
rotation number, $r_g$, and $D_n$ is the local diffusion coefficient. Since $|r_g-r'_n|\sim
q_n^{-2}$ and $D_n\sim q_n^{-5}$, one gets the estimate for the local exit time:
\be
\tau_n\sim q_n
\label{eq:01}
\ee
When $n\to\infty$ (i.e. when $K\to K_g$) one finds for the exit time, $\tau$, from the scale $n$
the following asymptotic behavior:
\be
\tau_n\simeq (1+r_g)^n = (1.618)^n
\label{eq:02}
\ee
To summarize, the following two facts \cite{chir2} constitute the basis of our
semi--phenomenological consideration:
\begin{itemize}
\item When approaching the chaos boundary from below, the average local exit time, $\tau_n$,
from the metastable island of the hierarchy scale $n$ is proportional to the number of periodic
orbits, $q_n$ (see \eq{eq:01}). This fact reflects the "local" structure of our phase space;
\item When $n\to\infty$, the local exit time, $\tau_n$, grows exponentially with the hierarchy
scale $n$ (see \eq{eq:02}). This fact reflects the "global" structure of our phase space and allows
to construct the estimate for the survival probability.
\end{itemize}

\section{Dynamics in the hierarchical tree--like phase space: diffusion on the boundary {\em vs}
diffusion in the bulk}

The hierarchical construction of self--similar sets implies that new smaller domains being properly
magnified (rescaled) with the limiting scaling factor $s_g$, coincide (in the statistical sense)
with the former ("parent") domains. This construction refers implicitly to the tree--like geometry.
The tree--like hierarchical geometry of the phase space of the standard mapping has been proposed
in the seminal analytic works \cite{meiss1,meiss2}. The authors have supposed that the states of
the system are the regions bounded by the low--flux cantori. Each state consists of an infinite
hierarchy of low--flux cantori of smaller scale. Between any two adjacent cantori there are
possible many other sub--hierarchies (or "island chains" in the terminology of \cite{meiss1}).
However only one such "island chain" in each hierarchical level is considered. Thus, the topology
of the full phase space can be represented as a following diagrammatic hierarchy, where the
hierarchy depth (level) is labelled by the same index $n$, appeared already in \eq{eq:0} as the
cutoff in the continuous fraction expansion:

\begin{figure}[ht]
\epsfig{file=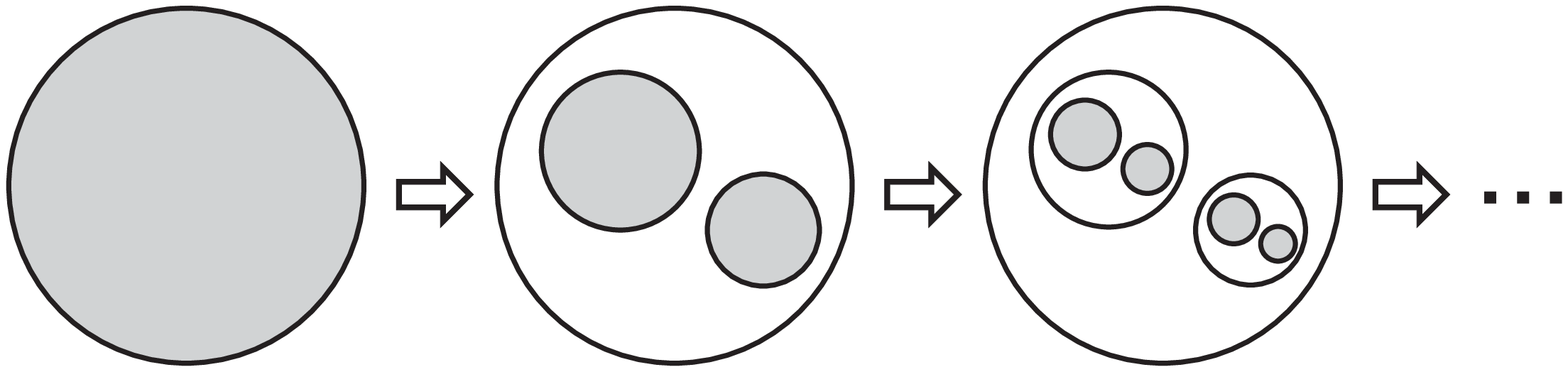, width=7cm}
\end{figure}

Knowing the local transition rates between neighboring states from microscopic computations, the
authors in \cite{meiss1,meiss2} have considered the random walk on the 3--branching tree and have
derived the corresponding infinite hierarchy of recursive relations for the survival probability.
The plausible conjectures about the closure of this hierarchy and subsequent analysis of the
dominant contribution to $P(t)$ allowed to extract the decay exponent $\nu\approx 1.96$ in
\cite{meiss1}. However this value is still rather far from numerically obtained exponent
$\nu\approx 1.4 \div 1.5$. Below we formulate an alternative point of view on dynamics on
hierarchical landscapes enabling to find the value of $\nu$ much closer to known numerical value of
the exponent $\nu$.

The hierarchical description sketched above is fully consistent with the consideration of a
"diffusion in a mountain landscape" \cite{hoffmann} appeared in a generic description of
"diffusion in hierarchies" regardless the subject of the chaotic Hamiltonian dynamics. For
heuristic consideration, let two sequences of real numbers, $A_n$ and $V_n$, be
correspondingly the sizes of landscape valleys (basins) and the heights of passages (energy
barriers) between these valleys with respect to some reference (energy) level. The basin sizes
and the barrier heights can be introduced iteratively. Suppose that some dynamical system is
located in a basin $A_0$ at the initial time moment. During the time $t_1$ the system
overpasses the lowest available height $V_1$ and the probability to find the system in some
state distributed initially on $A_0$, relaxes to the larger basin $A_1\supset A_0$.
Inductively, if the probability to find the system is located in the basin $A_n$ at time
$t_n$, then at time $t_{n+1}$ it relaxes into a larger domain $A_{n+1}\supset A_n$ by
surmounting the lowest available height $V_n$. Since by construction the $n$'s basin
hierarchically includes all basins $A_{n-1} \supset A_{n-2} \supset...\supset A_0$, there is
no difference between the waiting time in the $n$'s basin and the total time from the
beginning of the dynamical process. Thus, the survival probability under diffusion on
hierarchies depends essentially only on scaling of basin sizes, $A_n$, and barrier heights,
$V_n$, and exactly this fact allows one to use the tree--like spaces with alternative
(Archimedean or non--Archimedean) metrics (see, for example, \cite{PSAA}, \cite{OS}).

Now we are in position to describe the main idea of the present work. In our description the state
of the system for some value of $K$ near the chaos border is uniquely characterized by the number
of quasiperiodic orbits, $q_n$ at the hierarchy level $n$. Thus, the states are parameterized by
the Fibonacci numbers, $q_n$.

We consider the dynamics of the system as the transitions between different states for {\em given}
value of $n$. This is the key difference with the former description of \cite{meiss1,meiss2}
schematically outlined above. Namely, in the former approach the authors have considered the local
random walk in the {\em bulk} of the 3--branching Cayley tree with the transitions between
neighboring states belonging to two neighboring levels of hierarchy. Such a construction suggests
an Archimedean metric on the tree--like space. To the contrary, we propose to consider an effective
Markov dynamics on the {\em boundary} of the 3--branching Cayley tree truncated at the hierarchy
level $n$ ($n\gg 1$). We allow for {\em long distance} jumps which appear with the probability
prescribed by the limiting scaling factor $s_g$ {\em only between the states of the same hierarchy
level} $n$. This construction implicitly suggests non--Archimedean (ultrametric) space of states.

To be precise, our Markov process takes place in an effective "energy landscape" constructed in the
following way. We consider the phase space up to the scale $n$ meaning that we regard the low--flux
cantori (metastable islands) of scale $n$ as local (possibly degenerated) minima of an energy
landscape. Introduce now the selfsimilar scale--invariant structure of the {\em basins} of local
minima hierarchically embedded into each others. Namely, each larger basin of minima consists of
smaller basins, each of these consists of even smaller ones {\em etc}. Since the hierarchy level
$n$ is chosen arbitrary, each local minimum (i.e. low--flux cantorus of scale $n$) could be (and
should be) understood as a basin as well, containing again the scale--invariant hierarchy
("subtree") of basins of smaller scale. Such a hierarchy does not conserve the structure of phase
space as the islands of stability in the chaotic sea, but preserves the tree--like factorization of
islands of smaller scale out of the islands of larger scale as $K \nearrow K_g$. Note that in the
common description of tree--like factorization of low--flux cantori, the value of $n$ is counted
from the root of the tree, meaning that larger and larger values of $n$ correspond to metastable
islands of smaller and smaller scales.

To specify our long--distance--jump Markov process in terms of the transition rates between the
states (local minima), we construct the hierarchy of barriers between the basins of minima. Namely,
larger basins are separated by higher barriers, while embedded smaller basins are separated by
lower barriers. The Cayley tree can be regarded as a hierarchical "skeleton" of energy landscape
(but not as a space of states as in \cite{meiss1,meiss2}): the bulk vertices of the tree
parameterize the hierarchy of basins of local minima hierarchically embedded into each others; the
same vertices parameterize the barriers separating the basins.

The crucial requirement to our Markov process is as follows: the transition probability (per
time unit) between any two local minima is determined by the maximal barrier separating these
minima. This requirement is equivalent to the strong triangle inequality. Thus, basin sizes
and barrier heights can be expressed in terms of ultrametric distances between local minima.

In conventional description of ultrametric spaces accepted in applications of $p$--adic
mathematical analysis \cite{vldimirov} to dynamics on energy landscapess, the ultrametric distances
between local minima are labeled in such a way that smaller transition rates  (i.e. higher
barriers) correspond to larger ultrametric distances. Therefore, while specifying ultrametric
distances similar to the $p$--adic norm, $p^\gamma, \gamma=1,2,...,n$ ($p$ is the prime number), we
define the index $\gamma$ which numerates the hierarchy levels of Cayley tree in the direction
opposite to the one of index $n$. In this construction $\gamma=1,2,...n$ is counted from the {\em
boundary} of the tree up to the tree origin, namely, there is a hierarchy of ultrametric distances
between the states of scale $n$ possessing the values $p^1, p^2,...,p^n$.

The hierarchical landscape with growing barriers corresponds to the fact that the average
local exit time $\tau_n$ from the states of scale $n$ grows with $n$:  larger the tree
(i.e. deeper the hierarchy), longer the exit time averaged over all states on the tree
boundary. This condition does not yet defines completely our energy landscape. We should
satisfy another important requirement, viz, the average local exit time $\tau_n$ from the
states of scale $n$ grows as
\be
\tau_n\sim (q_n/q_{n-1})^n \simeq s_g^n
\label{eq:tau}
\ee
for $n\to\infty$, where $q_n$ is the $n$'s Fibonacci number and $s_g\approx 1.618$. This means
that the energy landscape, being expressed in terms of ultrametric distances between the
states of scale $n$, meets a specific behavior of the average local exit time, $\tau_n$, with
$n$, and the survival probability as well, through the relation to the Fibonacci numbers. This
is another key ingredient of our approach.

\section{The tree--like geometry and the Fibonacci numbers}

The procedure of construction of such a hierarchical landscape is described below. Anyway, to be
specific, in what follows we shall always keep in mind the 3--branching Cayley tree, ${\cal C}$, as
an example of an ultrametric space.

\begin{figure}[ht]
\epsfig{file=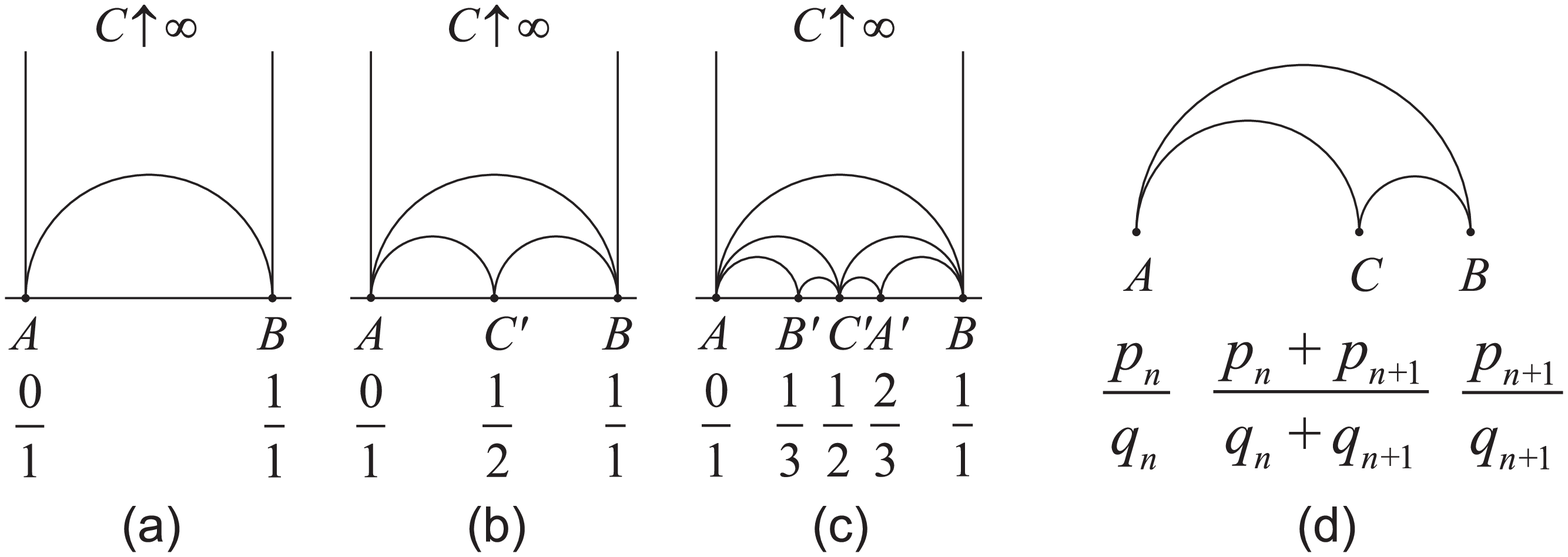,width=8cm} \caption{Inversions of zero--angled triangles: a), b), c)
-- three subsequent stages of tessellation of the half--strip ${\rm Im}\,z>0;\, 0\le{\rm Re}\,
z\le 1$ by the images of the triangle $ABC$; d) -- composition rule for the coordinates of the
vertices.}
\label{fig:1}
\end{figure}

The Fibonacci numbers have natural relation to the ultrametric geometry since they are connected to
some discrete symmetries of the hyperbolic space. In order to substantiate our construction, it
seems to be instructive to demonstrate briefly how the Fibonacci numbers appear in ultrametric
geometry and how they are connected to a 3--branching Cayley tree. Take the upper complex
half--plane $z=x+iy>0$ and consider the zero--angled curvilinear triangle $ABC$ bounded by two
vertical lines $AC$, $BC$ and a semi--circle $AB$ leaned on the real axis as shown in the figure
\fig{fig:1}a. Tessellate now the upper half--plane strip ${\rm Im}\,z>0;\, 0\le{\rm Re}\, z\le 1$
by the images of the triangle $ABC$. Two subsequent steps are shown in \fig{fig:1}b,c. These images
are obtained by sequential inversions (fractional--linear transformations) of the initial triangle
$ABC$. The images of all vertices of the triangle $ABC$ lie on the real axis ${\rm Im}\,z=0$ and
the coordinates of the vertices of neighboring triangles satisfy the following composition rule
\be
x_{n+2}= \frac{p_{n+2}}{q_{n+2}} \equiv \frac{p_n}{q_n} \oplus \frac{p_{n+1}}{q_{n+1}}
\stackrel{\rm def}{=} \frac{p_n+p_{n+1}}{q_n+q_{n+1}}
\label{eq:2}
\ee
shown in \fig{fig:1}d.

\begin{figure}[ht]
\epsfig{file=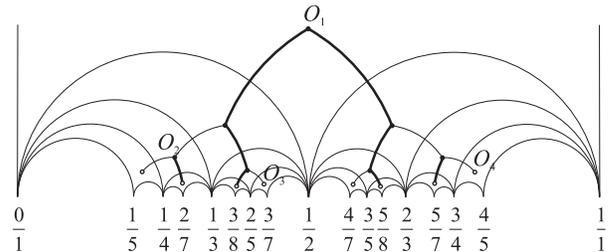,width=8cm} \caption{Few subsequent inversions of the zero--angled
triangle in the strip ${\rm Im}\,z>0;\, 0\le{\rm Re}\, z\le 1$ is shown). Inversions,
corresponding to Fibonacci sequences are shown in boldface.}
\label{fig:2}
\end{figure}

Define now the 3--branching Cayley tree isometrically embedded in the upper half--plane strip
${\rm Im}\,z>0;\, 0\le{\rm Re}\, z\le 1$ by connecting the centers of neighboring images of
zero--angled triangles by arcs being parts of semicircles leaned against the real axis -- see
the \fig{fig:2}. Recall that the embedding of a Cayley tree ${\cal C}$ into the metric space
is called "isometric" if ${\cal C}$ covers that space, preserving all angles and distances.
For example, the rectangular lattice isometrically covers the Euclidean plane $E\{x,y\}$ with
the flat metric $ds_E^2=dx^2+dy^2$. In the same way the Cayley tree ${\cal C}$ isometrically
covers the surface of the constant negative curvature, ${\cal H}$. One of possible
representations of ${\cal H}$, known as a Poincar\'e model, is the upper half--plane ${\rm
Im}\,z>0$ of the complex plane $z=x+iy$ endowed with the metric $ds_{\cal
H}^2=(dx^2+dy^2)/y^2$ of constant negative curvature. The composition rule \eq{eq:2} defines
the coordinates of corresponding triangle vertices, $\{x_n\}$. Hence, the rule (3) defines a
set of rational numbers parameterizing all bulk vertices of the Cayley tree of $n$
hierarchical levels. The Fibonacci sequences appear as the subsets of $\{x_n\}$ corresponding
to alternating left--right (or symmetric right--left) sequences of reflections of triangles
(these sequences are marked in \fig{fig:2} in boldface). For example, two "zigzags" starting
from the point $O_1$ correspond to two "principal" Fibonacci sequences, $f^{+}(O_1)=[1111...]$
and $f^{-}(O_1)=1-[1111...]$. The "secondary" zigzags, $f(O_2)$, $f(O_3)$ and $f(O_4)$
starting from the points $O_2$, $O_3$ and $O_4$ correspondingly, have the continued fraction
expansions: $f(O_2)=1-[121111...]$, $f(O_3)=[112111...]$ and $f(O_3)=[131111...]$.
Generically, all "zigzags" have the following continued fraction representation:
\be
f(\rm zig) = \left\{ \begin{array}{ll} [1\, m_2\, m_3\,...\, m_k\, 1111...] & \mbox{for
$1/2<x<1$} \medskip \\ 1-[1\, m_2\, m_3\,...\, m_k\, 1111...] & \mbox{for $0<x<1/2$}
\end{array} \right.
\label{eq:3}
\ee
with $k$ first arbitrary numbers $m_2,\, m_3,\, ..., m_k\; (m_1=1)$ are followed by the
"Fibonacci tail" of "1".

\begin{figure}[ht]
\epsfig{file=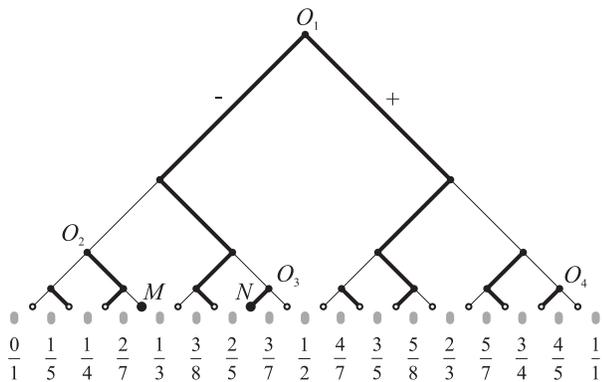,width=8cm} \caption{Topological structure of the graph obtained by the
successive applications of fractional--linear transformations in \fig{fig:2}. "Principal" and
"secondary" Fibonacci sequences are marked in boldface (see the explanations in the text).}
\label{fig:3}
\end{figure}

Washing out the metric structure of the set of fractional--linear transformations depicted in
\fig{fig:2} and leaving topological structure of the corresponding "zigzags" on the Cayley tree we
arrive at the relation between the Fibonacci numbers and an ultrametric space shown in \fig{fig:3}.

It should be emphasized that the ultrametric graph in \fig{fig:3} designates schematically the
hierarchically organized transition rates (energy barriers) between the states on the tree
boundary, while the bottom gray points labeled by rational numbers, $p_n/q_n$, parameterize
the bulk vertices of the ultrametric tree, and hence, they correspond to the transition rates
over the barriers. Again, the dynamics occurs only at this boundary, but not in the bulk of
the tree.

\section{Survival probability}

Let us come back to the computation of the survival probability, $P(t)$. Our main conjecture is as
follows. Extend the set of states of the dynamic system and suppose that all fractions $p_n/q_n$
parameterize some quasiperiodic orbits. The transition rates between any two different states $M$
and $N$ are defined by the relative "distance" between these states measured in number of
successive reflections (as in \fig{fig:2}) necessary to superpose the points $M$ and $N$. One might
thought about our system in the following terms. Take a 3--branching Cayley tree up to some
hierarchical level (generation) $n$ shown in \fig{fig:3}. Parameterize all the bulk vertices of the
tree by the sets of rational numbers $p_n/q_n$ following the reflection described above, and
consider the "ultrametric diffusion" on the boundary -- a Markov process with the transition rates
encoded in block-hierarchical kinetic matrix, known as "Parisi transition matrix" -- see \cite{OS},
\cite{avet1}. Thus, our jump--like Markov process is defined by the Parisi kinetic matrix whose
matrix elements are encoded by the set of rational numbers $p_n/q_n$.

The survival probability, $P(t)$, is the probability to find the system in the initial state
after $t$ jumps on the boundary of the Cayley tree (with hierarchically organized transition
probabilities). According to the works \cite{avetisov} (see also \cite{avet-nech} for the
transparent geometrical interpretation), the function $P(t)$ consists of additive
contributions from all possible directed paths on the $p$--adic tree (recall that in our case
$p=2$):
\be
P(t,\Gamma) =(p-1)\sum_{\{\gamma,j\}} p^{-\gamma}\, e^{\lambda_{\gamma,j} t} +p^{-\Gamma}
\label{eq:4}
\ee
The indices $\gamma$ and $j$ label correspondingly the hierarchical level of the tree ($1\le
\gamma\le \Gamma$) and the specific point in the hierarchical level $\gamma$ ($1\le j \le
p^{\Gamma-\gamma}$). Pay attention that now $\gamma$ is counted from the boundary of the tree
towards the root point (see the discussion above). The eigenvalues $\lambda_{\gamma,j}$, are
defined via the following construction
\be
\lambda_{\gamma,j}=-p^{\gamma} q_{\gamma}^{(j)} - (1-p^{-1})
\underbrace{\sum_{\gamma'=\gamma+1}^{\Gamma} p^{\gamma'} q_{\gamma}^{(j')}}_{\Sigma}
\label{eq:5}
\ee
where the sum denoted by $\Sigma$ runs along the tree from some vertex point labelled by the pair
of indices $(\gamma,j)$ to the root point $O_1$, and $q_{\gamma}^{(j)}$ is the transition
probability corresponding to the state labelled by $(\gamma,j)$.

The whole variety of directed sequences running from the hierarchical level $\gamma=1$ to the root
point $O_1$ is bounded by two "limiting" trajectories "logarithmic" and "linear". The "logarithmic"
is $\sigma_{\rm log}=\{1/5, 1/4, 1/3, 1/2\}$, and the "linear", being the "principal" Fibonacci
one, is $\sigma_{\rm lin}=\{3/8, 2/5, 1/3, 1/2\}$. The denominators in the "logarithmic" sequence
grow linearly with $\gamma$, $q_{\gamma}=\gamma$, while the denominators of the "linear" sequence
grow exponentially, $q_{\gamma}\simeq (1.618)^\gamma$. The notations "logarithmic" and "linear"
come from the fact that $V(\gamma,j)\sim -\ln q_{\gamma}^{(j)}$ can be considered as the effective
dimensionless local height of the potential barrier in the point $(\gamma,j)$. The "logarithmic"
landscape is associated with the "logarithmic" sequence, for which one has $V(\gamma)\sim \ln
\gamma$, while the "linear" landscape is associated with the "linear" sequence, for which one has
$V(\gamma)\sim \gamma$.

Taking into account that: i) the eigenvalues \eq{eq:5} of the transition matrix are given by
the weighted sums along different directed paths on the Cayley tree, and ii) the survival
probability does not depend on the multiplicity of paths on the tree with the same sequence of
transition probabilities, we can directly use the results of \cite{avetisov} for survival
probabilities in logarithmic and linear landscapes.

The survival probability for logarithmic and linear landscapes reads (see \cite{avetisov})
\be
P(t) \simeq \left\{\begin{array}{ll} \disp e^{-t/\ln 2} & \mbox{for $\disp q_{\gamma}=2^{-\gamma}\,
\gamma^{-1}$ (logarithmic)} \medskip \\ \disp C\; t^{-1/\alpha} & \mbox{for $\disp
q_{\gamma}=2^{-\gamma}\, 2^{-\alpha \gamma}$ (linear)}
\end{array} \right.
\label{eq:6}
\ee
where $c=\Gamma\left(\frac{1}{\alpha}+1\right) \left(-\Gamma_{p=2}(-\alpha)\right)^{-1/\alpha}$ and
$\Gamma_{p=2}(...)$ is the $p=2$--adic $\Gamma$--function (see \cite{avetisov} for details).

The appearance of the factor $2^{-\gamma}$ in the transition probabilities $q_\gamma$ in \eq{eq:6}
should be clarified. By definition, $\gamma^{-1}$ (for the logarithmic landscape) and $2^{-\alpha
\gamma}$ (for the linear landscape) are the transition probabilities over the barrier of the
hierarchy level $\gamma$ separating two basins. To get the transition probabilities between the
{\em specific points} $x$ (located in one basin) and $y$ (located in the second basin)---just these
probabilities enter in the kinetic equation on ultrametric trees---one should divide the transition
probability over the barrier by the number of states in the basin, which is in our case
$2^{\gamma}$.

The value of $\alpha$ can now be found straightforwardly. Rewrite the transition rate
$q_{\gamma}$ as
\be
q_{\gamma}=(1.618)^{-\gamma}=2^{-\alpha\gamma}
\label{eq:7}
\ee
From \eq{eq:7} one gets
\be
\alpha=\frac{\ln 1.618}{\ln 2}\approx 0.6942...
\label{eq:8}
\ee
One sees from \eq{eq:6} that in the large--time limit ($t\to\infty$) the decay of the survival
probability is exponentially fast on logarithmic landscapes. Since the survival probability
consists of additive contributions from all possible directed paths on the $3$--adic tree, only the
linear landscapes, i.e. the "principal" Fibonacci sequences, give the major contribution to the
survival probability in the large--time limit ($t\to\infty$), leading to the following algebraic
decay:
\be
P(t)\sim t^{-\nu} = t^{-\ln 2/\ln s_g } \approx t^{-\ln 2/\ln 1.618}= t^{-1.44}
\label{eq:9}
\ee
where $s_g=1+r_g = (\sqrt{5}+1)/2\approx 1.618$. Thus, we have $\nu \approx 1.44$.

\section{Conclusion}

Despite our value $\nu \approx 1.44$ is much closer to numerically obtained critical exponent
$\nu\approx 1.4\div 1.5$ than many other values of $\nu$ found in former analytic approaches
(except the one found in \cite{ketzm} and \cite{ven1,ven2}), we are far from a naive thought that
our simple consideration resolves the problem of analytic computation of the survival probability
in the nonlinear dynamical equation \eq{eq:rec} in some region near the chaos border. We only have
reformulated some particular problems of chaotic Hamiltonian dynamics in terms of Markov dynamics
on boundaries of ultrametric trees with transition probabilities prescribed by internal dynamics of
the system. Recall the two main ingredients of our consideration borrowed from \cite{chir2}: i) for
$K\nearrow K_g$ the average local exit time, $\tau_n$, from the metastable island of the hierarchy
scale $n$ is proportional to the number of periodic orbits, $q_n$; ii) for $n\to\infty$ the local
exit time, $\tau_n$, grows exponentially with the hierarchy scale $n$.

Note that our description does not contradict with the value $\nu=3$ found in scaling analysis
\cite{chir2} just at the boundary $K=K_g$. Namely our consideration "smears" the scaling $\tau_n
\simeq s_g^n$ to some region below $K_g$ where the "secondary" Fibonacci sequences starting from
the points $O_2$, $O_3$, $O_4$ in the \fig{fig:3} come into the play.

Let us end up this letter by saying that the conjectured approach offers a possibility to rise some
interesting questions concerning the internal structure of the standard mapping \eq{eq:rec}. For
example, it would be desirable to check numerically the existence of the "secondary" Fibonacci
sequences starting from the points $O_2$, $O_3$, $O_4$ in the \fig{fig:3}. In case of their
presence, one would be intriguing to think about the "phyllotaxis" \cite{douady,levitov} in chaotic
dynamical systems. By this conjecture we would like to attract the attention of researchers working
in nonlinear dynamical and chaotic systems to the language \cite{vldimirov} developed for the
description of stochastic processes in ultrametric spaces.

\begin{acknowledgments}

The authors are indebted to D. Shepelyansky for motivating us to touch this particular problem and
for valuable critical comments on all stages of our work. The work is partially supported by the
RFBR grants Nos. 07-02-00612a, 09-01-12161-ofi-m.  S.N. is grateful to Y. Fyodorov for  his idea to
apply the language of hierarchical matrices in Hamiltonian dynamics and to D. Ullmo for
introductory comments into the subject.

\end{acknowledgments}

\end{document}